\documentclass[
aps,%
12pt,%
final,%
notitlepage,%
oneside,%
onecolumn,%
nobibnotes,%
nofootinbib,%
superscriptaddress,%
showpacs]%
{revtex4}
\usepackage{graphicx}
\usepackage{amsmath}
\usepackage{amsfonts}
\usepackage{bbm}
\begin{document}
\title{On a microscopic representation of spacetime\protect{\footnote{The 
author thanks the Alexander von Humboldt-Foundation (Bonn, Germany) for 
financial support.}}}
\author{\firstname{Rolf}~\surname{Dahm}}
\affiliation{Permanent addr.: beratung f{\"{u}}r IS, G{\"{a}}rtnergasse 1, D-55116 Mainz, Germany}
\email{dahm@rolf-dahm.de}

%
\begin{abstract}
We start from a noncompact Lie algebra isomorphic to the Dirac
algebra and relate this Lie algebra in a brief review to low 
energy hadron physics described by the compact group SU(4). This
step permits an overall {\it physical} identification of the operator 
actions. Then we discuss the geometrical origin of this noncompact
Lie algebra and 'reduce' the geometry in order to introduce in
each of these steps coordinate definitions which can be related
to an algebraic representation in terms of the spontaneous symmetry
breakdown along the Lie algebra chain su*(4) $\longrightarrow$ 
usp(4) $\longrightarrow$ su(2)$\times$u(1). Standard techniques
of Lie algebra decomposition(s) as well as the (physical) operator 
identification give rise to interesting physical aspects and lead 
to a rank-1 Riemannian space which provides an analytic representation
and leads to a 5-dimensional hyperbolic space $H_{5}$ with SO(5,1) 
isometries. The action of the (compact) symplectic group decomposes 
this (globally) hyperbolic space into $H_{2}\oplus H_{3}$ with 
SO(2,1) and SO(3,1) isometries, respectively, which we relate to 
electromagnetic (dynamically broken SU(2) isospin) and Lorentz 
transformations. Last not least, we attribute this symmetry pattern 
to the algebraic representation of a projective geometry over the
division algebra $\mathbbm{H}$ and subsequent coordinate restrictions.
\end{abstract}
\maketitle

\section{Introduction}
This paper is intended as a conceptual paper focussing on the 
background of a spontaneously broken symmetry pattern, thus 
summarizing some of our previous results (\cite{dahm:2001},
\cite{dahm:2006}, \cite{dahm:2008}). This research originated
in effective descriptions of hadronic interactions \cite{dahm:diss},
however, instead of approaches like 'chiral perturbation theory'
or QCD we've chosen another way of realizing (and breaking) 
hadronic symmetries thus avoiding wellknown deficits of the 
standard approaches.
However, in order to avoid political and theological discussions
on these subjects in what one has {\it to believe}, we go back 
to a very simple ansatz by using nothing but the standard 
(15-dimensional) Dirac algebra which is common to all 'quantum'
approaches. And because the Dirac algebra is the very foundation 
of calculations in quantum field theory (QFT) and gauge theories, 
an identification of the operators of this 15-dimensional algebra 
is of considerable physical interest to (quantum) gauge theories,
too.

In the subsequent sections, we try to develop the idea and the
concept stepwise while we report on some work which originated 
in hadron physics and which became more and more interesting
as we illuminated the underlying geometry. So in addition to 
some recently published calculations \cite{dahm:2008} it is
noteworthy to present the geometrical and conceptual background
which we understand as underlying the usual Dirac and QFT
description. It is interesting to see that (in agreement with
Klein's 'Erlanger Programm') we can use (Lie) group and algebra
theory to obtain a finite as well as an infinitesimal (algebraic) 
description of this geometry, and the affine and differential 
geometry used nowadays turns out to be a subsidiary concept of 
this scheme. Moreover, it seems that Weyl's separation of
'{\"a}ltere' from 'infinitesimale Geometrie' (see e.g.~\cite{weyl:1988}) 
has lead over the decades to a tilt in favour of pure technical 
issues in terms of affine concepts and gauge theories whereas 
some of the underlying superior geometrical concepts fell into 
oblivion (especially those going beyond just the first approach 
by a tangential space).

\section{SU(4) versus SU$*$(4)}
So in order to find an 'entry point' to the subsequent discussion
we summarize our ({\it physically motivated}) assumption(s) which 
serve as ansatz - the identification of the number space we are 
going to use and an assumption about relativistic symmetry and its 
breaking.

As such, we can observe in chiral hadron theories (\cite{dahm:diss}, 
\cite{dahm:mex}) that with higher energies, the symmetry breaking 
becomes larger and that the symmetry scheme becomes worse and worse.
If we cover chiral SU(2)$\times$SU(2) symmetry by the larger compact
group SU(4), we see that SU(4) has some interesting properties when 
compared to the spectrum \cite{dahm:diss}, and in describing axial 
symmetry properties and charges, however, the SU(4) multiplets are 
not realized as (Wigner-Weyl) supermultiplets in the spectrum. 
Nevertheless, counting multiplet members and spin-isospin quantum 
numbers in the spectrum, the SU(4) dimensions can be identified if 
we group observable (distributed) multiplets together \cite{dahm:diss}. 
Moreover, it is known from other observations e.g.~in nuclear 
physics that SU(4) can serve as a good low energy symmetry, although
it is broken for higher energies.

So instead of writing the (nonlinear) chiral transformation in terms
of (complexified) Pauli matrices (which is a certain representation 
but mathematically and physically misleading) we choose a description 
in terms of real quaternions right from the beginning. Yes, at a first 
glance this seems to be artificial, however, we benefit twice from 
the fact that the (real) quaternions constitute a division algebra: 
mathematically in that we have well-defined inverses throughout our 
calculations, and we can use a well-defined and consistent division 
when working with the real quaternions, thus avoiding some magic 
(commuting) '$i$'s which otherwise occur at various places within 
the calculations. This obvious benefit allows to treat nonlinear 
chiral transformations {\it completely} instead of calculating with 
certain awkward 'expansions' and/or weakly justified 'power series'. 
The much larger advantage, however, is an identification of usual 
chiral transformations as subsets of quaternionic M{\"o}bius 
transformations which of course suggests to study a projective 
quaternionic geometry and start from scratch with geometry. 
Physically we benefit in that this identification is much closer to 
the {\it physical measurement process} and much more evident in that 
we compare a certain measure (respectively, an observable) with a 
known unit measure at various distances, i.e. we need unique inverses 
to compare with the unit by {\it dividing} the unit, and we need the 
full tool set of a projective geometry, too, to transport and scale 
the observations appropriately.

This simple reasoning leads to the symmetry group Sl(2,$\mathbbm{H}$)
which one can embed into complex spaces using appropriate coordinate
sets and 2$\times$2 matrix representations (see e.g. \cite{helgason:1978} 
or \cite{gilmore}) so that we obtain the isomorphic noncompact symmetry 
group SU$*$(4) on complex representation spaces. Simple calculations 
show that (up to a misplaced and additional commuting '$i$' in the usual 
definition) the operators of the Dirac algebra are isomorphic to the 
Lie algebra generators su$*$(4). Hence it is self-evident to understand
the occurence of an SU(4) symmetry at low energies as an approximation 
respectively as the remnant of the original relativistic symmetry SU$*$(4).
Following this reasoning, the 'difference' in the Lie algebras su(4) 
and su$*$(4) has to be responsible for dynamics whereas the (compact) 
part common to both su(4) and su$*$(4) should generate {\it dynamically} 
further observable symmetries (as isotropy group). Last not least, we can 
use the spin-isospin construction of SU(4) with its {\it physical operator 
identification} as well as the state identification (see \cite{dahm:diss}, 
\cite{dahm:mex}) in order to check the physical significance of our 
calculations and results.

The technical machinery to investigate such issues more precisely is 
well-known from Lie algebra theory, and we find a compact subgroup USp(4) 
of dimension 10 of both SU(4) and SU$*$(4) as well as a 5-dimensional 
operator set $p$ mapped by the exponential onto a coset space $\exp(p)$ 
which we can identify with SU$*$(4)/USp(4). So we find a spontaneously 
broken symmetry with a (local) USp(4) symmetry group (see also 
\cite{dahm:2006} and \cite{dahm:2008}). This is an irreducible Riemannian 
globally symmetric space (type AII, \cite{helgason:1978}) which we can 
describe technically in terms of a nonlinear sigma model with appropriate 
representations (or by differential geometry/ fiber bundles with local 
USp(4) symmetry or induced representations, resp.). Here, however, we 
postpone the technical details and focus instead on the geometrical 
background in terms of '{\"a}ltere Geometrie'. Why? First of all, 
we have to identify and define suitable coordinates and appropriate 
coordinate (transformation) rules which we can use {\it in a second 
step} to perform calculations, i.e.~they serve as an appropriate 
(algebraic) representation of this geometry. It is only the very 
{\it last step} that we need to use affine coordinates and concepts, 
which then results in 'local' coordinate systems with emphasized 
points (or special (restricted) transformation structures), or which
results in certain representation(s) for objects to which we then 
attribute physical properties or behaviour and where we can apply 
certain optimization (or action) principles like Lagrangian or (by
using 'more suitable' coordinate definitions) Hamiltonian formalisms 
to extract (infinitesimal) properties in terms of equations of motion. 
We will see that we can short-circuit some of these typical standard 
mechanisms by using geometry to understand the vacuum structure and 
write down the geodesics to determine the physical behaviour so that 
infinitesimal equations provide no more information than the finite 
representation.

\section{Geometry}
\begin{figure}[ht]
\includegraphics[scale=0.6]{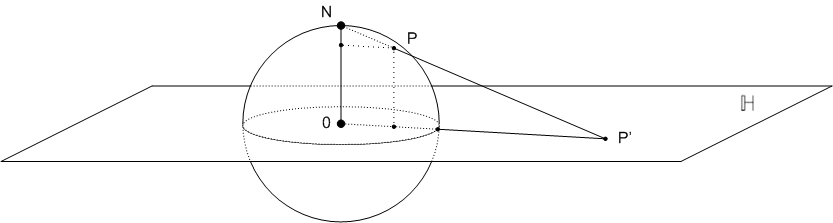}
\caption{Projection \protect{$S^{4}$} onto \protect{$\mathbbm{H}$}}
\label{fig:stpr}
\end{figure}

In some preceding publications, we've already presented the picture
given in figure \ref{fig:stpr} which can be seen in analogy to the
Riemannian case based on the division algebra $\mathbbm{C}$.
However, the figure in the given form already anticipates some 
preceding and important steps which are already incorporated in the 
given definition of points and axes but which otherwise would allow 
for additional degrees of freedom. So at next, we follow the reasoning 
and the construction process of the projective line and appropriate 
coordinate systems as discussed in \cite{klein:1928} for the (commutative) 
division algebra(s) of real (and complex) numbers.

\subsection{Projective Geometry and Coordinates}
Although at a first glance projective geometry is related to 
points, lines, connections of points and sections of lines (incidences),
in order to establish duality it is possible to use polar relationships 
\cite{klein:1928}, i.e. to study conic sections in the 'plane' $\mathbbm{H}$
by various methods.

Having to include the projection point in the plane at infinity
(which maps to $N$), it is helpful to introduce projective coordinates 
(or homogeneous coordinates if we fix the fundamental points at 0 and 
$\infty$) in the projective line by a set $(\rho q_{1}, \rho q_{2})$, 
$\rho\neq 0\in \mathbbm{H}$. Already here we see that noncommutativity
of quaternions gives rise to an additional Gl(1,$\mathbbm{H}$) 
transformation of the affine coordinates (depending on the definition 
of division) as $\rho^{-1}$ exists in $\mathbbm{H}$. If now in addition 
to the two fundamental points we {\it define} an unit point $\mathbbm{1}$,
$q_{1}=q_{2}$, we have three points which fix the coordinate system 
and we can use the cross-ratio ('Doppelverh{\"a}ltnis') to introduce 
measures, metric structures and more with respect to each fourth 
(quaternionic) point, respectively. As an immediate consequence, we 
can introduce real measures and coordinates by suitable cross-ratios 
(and geometrical configurations), we can express distances and angles 
naturally by logarithms of certain cross-ratios, and we end up with 
metric properties which we can control and whose background we 
understand {\it geometrically}. The projective line offers automatically 
the typical $4\pi$-behaviour of spinorial reps because the line is 
closed by one 'infinite point'. Moreover, the cross-ratio allows 
straightforward access to invariants of four points under projective 
transformations of the line or to its value in two projective coordinate 
systems, respectively, whereas $\exp$ and $\log$ reflect in group and
algebra theory. The discussion of complex numbers in the plane in 
relation to real parameters of $S^{2}$ as well as their occurence 
in cross-ratios \cite{klein:1928} can be (carefully) generalized to 
$S^{4}$, and we may as well discuss quaternionic and real results 
of cross-ratios as we find quaternionic transformation properties 
reflected in the corss-ratios. $S^{4}$ can also serve to define an 
orientation of a path between two point $p_{1}$ and $p_{2}$ of $S^{4}$, 
dependent on whether we pass the northpole on a great circle in the 
order ($p_{1}$, $p_{2}$, $\infty$) or ($p_{1}$, $\infty$, $p_{2}$). 
Last not least, using only projective methods (see \cite{klein:1928}, 
ch.~V,~\S 2 and \cite{klein:1928}, references) one can construct a 
complete coordinate system on the line. Following \cite{klein:1928} 
yields further interesting results, however, here we want to apply 
some of the aspects mentioned above with respect to our original 
physical problem by choosing twofold (homogeneous) quaternionic 
coordinates $q_{1}$ and $q_{2}$ and discussing Sl(2,$\mathbbm{H}$)
transformations\footnote{Internally, we use the shorthand notation 
'QPT' (quaternionic projective theory) to denote this framework.}.

\subsection{Sphere Rotations}
Related to the different coordinate types mentioned above, we can 
of course study the effects of sphere rotations. If we are thus to 
work with the two projective/homogeneous quaternionic coordinates 
$q_{1}$ and $q_{2}$ or the (affine) coordinate $q$ of the (finite)
quaternionic 'line' $\mathbbm{H}$, respectively, we can introduce 
(see e.g.~\cite{dahm:diss}, \cite{dahm:2001} or \cite{dahm:2008}) 
a 'spinor' $\psi$ by $\left(\begin{smallmatrix}q_{1} \\ q_{2}\end{smallmatrix}\right)$ 
and express the quaternionic M{\"o}bius transformations $f(q)$ by 
a mapping onto $2\times 2$ quaternionic matrices acting on the two 
homogeneous coordinates. The associated 
group is Sl(2,$\mathbbm{H}$) but we can restrict these M{\"o}bius 
transformations of course to subgroups of Sl(2,$\mathbbm{H}$), or 
when represented on complex spaces, to subgroups of SU$*$(4). 
Especially there, we can think of rotations 
keeping the 'norm' $\psi^{+}\psi$ invariant, which translates to 
the constraint $q_{1}^{+}q_{1}+q_{2}^{+}q_{2}=\mathrm{const}$ in 
terms of projective/homogeneous coordinates, and we naturally obtain 
the 'Hopf map' $q_{1}^{+}q_{1}+q_{2}^{+}q_{2}=1$, i.e. the map 
$S^{7}\longrightarrow S^{4}$. In our geometrical picture of the
original geometry, the sphere rotations correspond to unitary 
quaternionic transformations (i.e. U(2,$\mathbbm{H}$)) of the 
homogeneous coordinates which, when expressed in complex coordinates,
correspond to the compact group USp(4). So we've found a simple 
geometrical explanation of the maximal compact subgroup of our 
original picture, and we find conic sections in the projection
'plane'.

Moreover, because we can identify 'Dirac spinors' in this scenario
in various contexts\footnote{For example, to understand the (massive)
Dirac spinor components $u$ used in \protect{\cite{bjoedrell}}, we 
can introduce quaternionic spinor structures $\chi=(q, q^{+})^{T}$ 
respectively linear combinations $\chi'=(q\pm q^{+}, q\mp q^{+})^{T}$ 
with only four real dimensions but with well-defined 'conjugation' 
properties under quaternionic conjugation, or we may use an 
associated $2\times 2$ matrix representation acting on this 
'spinor' space of twofold homogeneous coordinates to exchange the
spinor components. However, here we postpone the details of this 
representation and the related discussion of QFT to an upcoming 
paper \protect{\cite{dahm:prep}}.} it is natural to find the Hopf 
map realized in QFT - this map it is a direct and straightforward 
consequence of the quaternionic M{\"o}bius transformations above
and a constraint to bind the two coordinates; because USp(4) is 
compact, we find a related conserved (unitary) norm in terms of 
the two homogeneous coordinates. The 'way back' to single quaternions
is possible due to the very existence of a division in $\mathbbm{H}$ 
since we may represent $q=q_{1}q_{2}^{-1}$, $q\in \mathbbm{H}$, 
and moreover, because we still have the freedom of an appropriate 
Gl(1,$\mathbbm{H}$) coordinate transformation by a real quaternion 
$\rho\neq 0$ as cited above. Last not least in this context it 
is important to remember the Cayley-Dickson construction of the 
division algebras where an additional complexification of the 
next lower division algebra may be represented by a skew-symmetric 
$2\times 2$ matrix 
$\left(\begin{smallmatrix} 0 & -1 \\ 1 & 0 \end{smallmatrix}\right)$. 
This construction scheme has important influence not only on 
the (overall) coordinate definition(s) but also highlights
certain matrix transformations in the groups as we will see 
with the operator\footnote{We use the definitions of 
$Q_{\alpha\beta}$ given in \protect{\cite{dahm:2008}}.} 
$Q_{02}$. USp(4) as a group which respects both an unitary and 
a symplectic (i.e.~mainly an orthogonal) norm benefits in an 
appropriate representation from a clear separation of conjugation
(of the field) and transposition of matrices and spinor reps 
(see also \cite{helgason:1978}). Later we'll discuss $Q_{02}$ 
as an U(1) generator as well as we need its discrete symmetry 
properties as presented in \cite{dahm:2008}, especially 
$Q_{02}^{2}=-1$.

\subsection{'{\"A}ltere Geometrie'}
Although we've presented above some geometry besides the 
frameworks used in today's models and calculations, we do 
not want to run (affine) coordinate approaches or purely
infinitesimal methods down. However, it appears necessary 
to treat some aspects of '{\"a}ltere Geometrie' at least 
on an equal footing as the 'infinitesimaler Zug'~\cite{weyl:1988},
just in order to understand more background of some current 
technical frameworks. '{\"A}ltere Geometrie' yields appropriate 
coordinate definitions and coordinate systems which allow
to go {\it beyond} affine models, it introduces naturally
and almost automatically cross-ratios, measures and their 
logarithmic dependence, metric structures as well as 'norm 
conserving' transformations related to properties of the 
cross-ratio and to polar relationships. Moreover, we can 
handle and understand fix points and invariants so that it
is possible to study certain conic sections at infinity, 
i.e. we can treat 'light cones' and 'vacua'. Last not least, 
we find geodesics, isometries and curvature from geometry,
and we can always proceed to certain suitable coordinate 
definitions to discuss differentiability or analyticity, 
respectively, because we can realize appropriate tangent 
structures and linear operators, even in differential 
representation (but not necessarily!).

In addition to the nonlinear (group based) discussion of 
sigma models, above we've tried to work out the correspondence 
with the geometrical reduction steps. So the 'geometrical 
chain' we've begun when discussing general transformations
of two independent homogeneous coordinates $q_{1}$ and $q_{2}$, 
which we've then restricted via a metric relation/a constraint 
$f(q_{1},q_{2})$ and which could be finally restricted to 
a single (affine) quaternion $q$ is reflected in the (matrix) 
representation of the transformation(s) SU$*$(4) $\longrightarrow$ 
SU$*$(4)/USp(4) $\longrightarrow$ USp(4)/Gl(1,$\mathbbm{H}$). 
This becomes more apparent if we look at the Lie algebras 
and remember the fact that Gl(1,$\mathbbm{H}$) is the 
covering group of U(2) as well as of SU(2)$\times$U(1), 
i.e. on the Lie algebra level repectively in infinitesimal 
models for the last reduction step we are discussing here 
nothing but a (local) representation of a (well-known) 
su(2)$\oplus$u(1) Lie algebra.

\section{Further Aspects and Outlook}
To present some physical consequences and results, we discuss 
some technical aspects. 

\subsection{Technical Aspects}
As we've already defined (see \cite{dahm:2008} and references 
therein) an operator basis $Q_{\alpha\beta}$ in terms of 
spin$\otimes$isospin operators, we can transform this 
representation (pointwise) to an alternate basis in terms 
of an isospin$\otimes$spin representation. To distinguish 
the two basis systems according to their different physical
content, we have introduced the operator set 
$\mathcal{Q}_{\alpha\beta}:=Q_{\beta\alpha}$ for the second 
representation. Choosing the representation of the Dirac 
algebra according to \cite{bjoedrell}, we find
$\gamma^{0}=i\mathcal{Q}_{30}$, $\gamma^{j}=-i\mathcal{Q}_{2j}$ 
and $\gamma^{5}=i\mathcal{Q}_{10}$, which differs from SU$*$(4) 
by the additional commutative $-i$ in the definition of 
$\gamma^{j}$. However, this arbitrary complexification can 
be traced back in literature to an at that time 'suitable' 
definition of the Lorentz metric i.e.~to the relative 
complexification of the coordinate derivatives as a 
'postulate'~\cite{dahm:diss}. Within SU$*$(4) however, 
{\it all} (relative) phases of the operators are of course 
fixed, and there is {\it no freedom} for arbitrary 
complexifications, especially not with {\it commuting} '$i$'s. 
Instead, within SU$*$(4) the skew-symmetric operator $Q_{02}$ 
(resp. $\mathcal{Q}_{20}$) plays an important r{\^o}le with 
respect to complexification according to the embedding of 
division algebras by 2$\times$2 matrix reps and the Caley-Dickson 
process.

As we've already mentioned above, we do not find supermultiplets 
in the low-energy regime of the spectrum. Instead, the mesons 
and fermions are pretty well grouped according to their
respective spin-content\footnote{Here, we cannot run the
discussion to what extent the spin grouping of hadrons in
the spectrum and the comparison of standard QFT calculations
to experimental data via {\it complex} pole analysis are 
related. For now, we take the grouping in spin multiplets 
for granted.} so in order to understand properties of the 
spectrum we have to break SU$*$(4) (resp. SU(4)) by an 
appropriate mechanism. Because we know of the symplectic 
symmetries of dynamical systems, it is reasonable to think 
of a decomposition of the global symmetry with respect to 
the maximal subgroup USp(4), and we are thus lead directly 
to a discussion of SU$*$(4)/USp(4) which is isomorphic to 
$H_{5}$ (see \cite{dahm:2008} and references) and has beautiful
(differential) properties as globally Riemannian symmetric
space. This breakdown can be treated and understood as a 
spontaneous breakdown of the global SU$*$(4) symmetry to a 
local symplectic symmetry which both are acting on the 
5-dimensional hyperbolic space $H_{5}$. The consequences of 
this approach are an USp(4) realization on the 5-dimensional 
space $H_{5}$ which itself can be either expressed in terms 
of hypercomplex numbers or embedded into $\mathbbm{R}^{6}$ 
by six (hyperbolic) coordinates which, of course, are {\it not} 
independent, and five Goldstone bosons related to shifts 
of the vacuum structure of the (local) symplectic symmetry 
group and it's 'origin' $\mathbbm{1}$ (resp. the origin $0$ 
in the Lie algebra).

In \cite{dahm:2008}, we have already discussed the necessary
algebraic tool sets (see also \cite{helgason:1978}) and we 
have given the isometry groups related to the coordinates
and coordinate sets of the coset space $\exp(p)$. These
isometry groups (SO(5,1) for $\exp(p)$ and SO(2,1) and 
SO(3,1) for the 'generator sets' $\{iQ_{01}, iQ_{03}\}$ and 
$\{Q_{j2}\}$, respectively) can be interpreted as coordinate
transformations of coordinate systems introduced within the 
respective generator sets (which themselves generate 
hyperbolic (sub-)spaces $H_{2}$ and $H_{3}$), so it is 
nothing but the hyperbolic structure (i.e.~the negative 
curvature) of the coset space which furnishes the dynamics 
with these three symmetry groups. The concept of spontaneous 
symmetry breakdown, i.e.~the concept of a {\it local} USp(4) 
symmetry, then requires to introduce coordinate systems with 
respect to each {\it new} vacuum definition in $\exp(p)$ so 
we have to change the respective coordinate representation 
as well\footnote{In other words, in su$*$(4) we may use 10
real parameters to determine the local usp(4) coordinate 
system whereas we can use 5 more real parameters to connect
and compare the respective USp(4) ground states or eliminate
superfluous components (see \cite{dahm:2001}, sect.~3 and
\cite{dahm:2006} with respect to the Dirac equation)}. Hence 
depending on the choice the new vacuum within $\exp(p)$, we 
find coordinate transformations either governed only by 
SO(2,1) or SO(3,1), or we have to consider connections in 
the 'full' coset space which are described by SO(5,1) 
covariance respectively by its double covering SU$*$(4). 
So already at this point it is evident that we can think 
of certain equivalence classes (and 'sub'vacua) in that we 
understand the various isometries as hyperbolic motions or 
we can use the gauge (and affine connection) concept as well 
for these hyperbolic spaces. As a consequence of hyperbolic 
coordinate representations in $\mathbbm{R}^{n}$, if we are 
going to associate the (physical) time with the 0-component 
in $H_{3}$ (i.e.~$\sim\cosh$), we naturally obtain strong 
causality ($\cosh > 0$, $\cosh > \sinh$). Moreover, the 
(physical) time is {\it not} independent from the (physical) 
space coordinates (i.e. $\sim\sinh$) (as we should know from 
Lorentz 4-space), but instead of formally six real independent 
coordinates (or four in the case of $H_{3}$), we have only 
{\it five} (or {\it three}, resp.) due to the hyperbolic 
constraint(s) on these {\it real} representation spaces. 
This, of course, has consequences for time and space within 
a real (pseudo-)orthogonal coordinate identification as being 
only a {\it derived} concept due to the representation of 
$H_{3}$ chosen on $\mathbbm{R}^{4}$. This carries forward 
to the differentials $\mathrm{d}x_{i}$ of these coordinates 
which are {\it not independent}, too. However, such problems 
can be circumvented by working with the underlying 
(noncommutative) Lie algebra or an appropriate hypercomplex 
representation. Nevertheless, we can of course always introduce 
local/affine coordinate sets as well as appropriate differential 
operators founding on these coordinates in order to realize at 
least the linear approximation (the tangent spaces) of the 
underlying geometry. Differential geometry provides additional 
tools sets (see \cite {helgason:1984}) in that due to the rank-1 
of $H_{5}$ one can realize the (more general) Laplace-Beltrami 
operator(s) $\mathcal{L}_{X}$ at a point $p$ of a rank-1 coset 
$X=G/K$ in terms of geodesic polar coordinates $(r,\theta_{i})$
according to
\[
\mathcal{L}_{X}=\frac{\partial^{2}}{\partial r^{2}}
+\frac{1}{A(r)}\frac{\mathrm{d}A}{\mathrm{d} r}\frac{\partial}{\partial r}
+\mathcal{L}_{S}
=\frac{\partial^{2}}{\partial r^{2}}
+\frac{1}{\sqrt{\overline{g}}}\frac{\partial\sqrt{\overline{g}}}{\partial r}
\frac{\partial}{\partial r}
+\frac{1}{\sqrt{\overline{g}}}\sum_{i,j=1}^{n-1}\frac{\partial}{\partial\theta_{j}}
\left(g_{ij}\sqrt{\overline{g}}\frac{\partial}{\partial\theta_{i}}\right)
\]
($\mathcal{L}_{S}$ is the Laplace-Beltrami operator on the 
sphere $S$ in $X$, $A$ the area of $S_{p}(r)$)  
so that for the hyperbolic spaces $H_{n}$ we obtain \cite{helgason:1984}
\begin{equation}
\mathcal{L}\,=\,\frac{\partial^{2}}{\partial r^{2}}\,+\,
(n-1)\,\coth r\frac{\partial}{\partial r}\,+\,
\left(\sinh r\right)^{-2}\,\mathcal{L}_{S}\,.
\label{eq:LBop}
\end{equation}
If we take the view of potential theory and spectral functions for 
the negatively curved rank-1 space, the potentials expected in the 
context of $H_{2}$ and $H_{3}$ should show the same (radial) behaviour 
$\sim r^{-1}$ if we take the space $H_{5}$ as a basis for the dynamics. 
This behaviour is interesting later after having identified $H_{2}$ 
and $H_{3}$ and the related physics.

With respect to \cite{dahm:2008}, sect.~3, we want to emphasize 
the possibility to decompose the 10-dimensional USp(4) generator 
set, i.e. the Lie algebra usp(4), further and define a CI-type 
sigma model. As we've set out, it is possible to understand this 
model in terms of a su(2)$\oplus$u(1) Lie algebra (whose generated 
local symmetry group can be covered by Gl(1,$\mathbbm{H}$)) and a 
$Q_{02}$-complexified (noncommutative) vector space of real dimension 
six. Hence. besides all the differential geometry related to this 
Hermitean symmetric space and the $Q_{02}$-complexification, we 
{\it may} define a norm on this space and a SU(3) automorphism 
group to rearrange these three (hyper-)complex coordinates while 
keeping the norm invariant. However, with this 'internal' SU(3) 
group, the remaining su(2)$\oplus$u(1) algebra governing the local 
symmetry as well as with the isometries discussed above (which are 
inherent already in the pure SU$*$(4) approach, i.e in the Dirac 
algebra!) one should take care when introducing additional unitary
symmetries by hand. {\it At least, it is necessary to think about 
double counting and/or the correct physical identification of the 
additional parameters and states}.

Here in the context of an SU$*$(4)/USp(4) realization, i.e.
with a global SU$*$(4) and a local USp(4) symmetry, it is now 
necessary to identify the vacuum (and its substructure(s)) 
as well as the transformations and the action of the group
on the coset (and thus the geodesics). Note that right from
the beginning, we have an overall consistent microscopic (and 
noncommutative) framework in terms of $Q_{\alpha\beta}$, and 
we are now trying to identify further physical processes by 
investigating the 'light cones'/vacua and geodesics. Put in terms of 
\cite{piranietal:1972}, if this succeeds we then know the 
local geometric structure of the manifold in that we know 
the conformal structure (the action of SU$*$(4) and the null 
cones), the projective structure (geodesics in $H_{5}$) as 
well as the affine structure, i.e. around each point of the 
manifold there exists an infinitesimal affine geometry 
(determined by 10+5 real parameters).

\subsection{Physical Consequences}
So in order to understand more of the underlying physics and 
the related identification of operators and transformations, 
we can benefit from our original starting point SU(4) with 
the thorough operator identification in the low energy-regime 
of the hadron spectrum, i.e. we know spin, isospin and axial 
transformations from our basic construction process~\cite{dahm:diss}.

Formally, we can act with the ten usp(4) generators on the 
coset space and see what happens. Hence if we decompose usp(4) 
into the three operator sets $Q_{02}$, $\{Q_{j0}\}$ and 
$\{iQ_{j1}, iQ_{k3}\}$ where $Q_{02}$ generates an U(1) symmetry, 
$\{Q_{j0}\}$ commute with $Q_{02}$ and generate (by themselves) 
an SU(2) symmetry, and $\{iQ_{j1}, iQ_{k3}\}$ generate the 
6-dimensional (type CI) coset space where the sets $\{iQ_{j1}\}$ 
and $\{iQ_{j3}\}$ are in addition related by a $Q_{02}$
multiplication, i.e. a relative (noncommutative) 
complexification~\cite{dahm:2008}.

If we define the (infinitesimal) action of $Q_{02}$ via
$\delta\,\cdot\, = [Q_{02}, \cdot\,]$, we can act on an element 
of $\exp p$. In this case we find that $\delta$ acts only on the 
$H_{2}$ subspace generated by $\{iQ_{01}, iQ_{03}\}$ wheras 
$\{Q_{j2}\}$ is invariant. We know already from \cite{dahm:2008} 
that $\{iQ_{01}, iQ_{03}\}$ is a Lie triple system (and a totally 
geodesic submanifold), and that $\{iQ_{01}, Q_{02}, iQ_{03}\}$ 
generate (noncompact) SO(2,1) isometries of this 2-dimensional 
hyperbolic space. If we remember that in our original SU(4) scheme
the compact generators $\{Q_{01}, Q_{02}, Q_{03}\}$ generate the 
isospin symmetry (see \cite{dahm:diss}, \cite{dahm:mex}), then 
the only and straightforward conclusion is that we {\it have to} 
associate $Q_{02}$ with the electromagnetic/electroweak U(1) 
symmetry transformation. Hence, in our coset space $H_{5}$ we 
identify two degrees of freedom $\{iQ_{01}, iQ_{03}\}$ on which 
the operator $Q_{02}$ acts by a U(1) rotation so we can try to 
associate them with two (real) polarization degrees of freedom. 
On the other hand, we know of the SO(2,1) isometry transformations 
of this coordinate set which we may equally well express in
terms of a SU(1,1) representation. This corresponds once more 
to the fact that $iQ_{01}$ and $iQ_{03}$ differ by a (noncomutative)
$Q_{02}$ multiplication. So we can rewrite the $H_{2}$-element
$a\,iQ_{01}+b\,iQ_{03} = iQ_{01}(a+b\,Q_{02}) = (a\,Q_{02}+b)\,iQ_{03} = (a-b\,Q_{02})iQ_{01} = \ldots$ 
equally well in terms of a (hyper-) complex coefficient related 
to only one remaining 'dimension'. This motivates using additional 
complex representations for masses and charges in terms of either 
a single complex number or using instead homogeneous complex 
coordinates by twofold spinor representation like in the Higgs 
(spinor) description.

However, the main result of this first step is the observation 
that compact isospin is broken to {\it noncompact} SO(2,1) by the 
spontaneaous breakdown of SU$*$(4)/USp(4) and that only an U(1) 
(as generated by $Q_{02}$) survives this breakdown. So there is 
{\it no} conserved isospin but we expect SO(2,1) to govern the 
spectral distribution instead of SU(2) isospin, especially at 
higher energies/masses. Nevertheless we have the (local) 
isometries SO(2,1) and the constraint via $H_{5}$ and its SO(5,1) 
isometries to 'rearrange' the coordinates in order 'to heal' 
and 'absorb' this isospin breaking a little bit\footnote{See 
also the discussion of isospin breaking in \cite{dahm:diss} with
respect to experimental data.}. The fact that $Q_{02}$ commutes 
with $\{Q_{j2}\}$, i.e. that $H_{3}$ is invariant under $Q_{02}$, 
allows to understand $H_{3}$ as a vacuum ('ground state') with 
respect to $Q_{02}$ and as bounded by an 'electromagnetic light 
cone', respectively. So we define
$\delta_{em}\,\cdot\, = [Q_{02}, \cdot\,]$.

The second compact generator set $\{Q_{j0}\}$ of usp(4) generators
shows similar behaviour when acting on $H_{5}$. This time, however, 
$H_{2}$ is invariant, i.e. $[Q_{j0},iQ_{01}]=[Q_{j0},iQ_{03}]=0$, 
so $H_{2}$ is bounded by a 'dynamic' light cone, but the $H_{3}$ 
subspace of $H_{5}$ is SU(2) rotated according to our original 
decomposition pattern from the su$*$(4) Lie algebra, i.e. 
$[h,p]\subset p$, and as a reasult we find another (rotated) 
element of $H_{3}$, $[Q_{j0},Q_{k2}]\sim Q_{l2}$. So we 
state the action of a compact transformation on a noncompact
(hyperbolic) space where the generators $\{Q_{k2}\}$ are a 
Lie triple system and $\{Q_{j0},Q_{k2}\}$ is isomorphic 
to the Lorentz algebra. Moreover, the commutator 
$[Q_{j2},Q_{k2}]\sim Q_{l0}$, i.e. $[p,p]\subset h$, 
according to $Q_{02}^{2}=-1\sim Q_{00}$ generates a compact 
(Wigner) rotation which is evident from the underlying su$*$(4)
algebra.

Accounting for these aspects, it is obvious to associate these
symmetry transformations with the Lorentz group and $H_{3}$ 
with spacetime. Hence formally (using \cite{GMS}), we are
finished with respect to the title of the paper because we
have related a microscopic representation originating from
QPT to $H_{3}$ and Lorentz transformations. So we can choose
directly an appropriate (real) representation of $H_{3}$ on 
$\mathbbm{R}^{4}$. In pursueing this idea, however, we may
identify points in $H_{3}$ with spacetime events, and we 
are lead to a hyperbolic geometry of points, geodesic lines, 
incidences, triangles, etc.) in $H_{3}$ which map to
$\mathbbm{R}^{4}$. This is no new result at all because e.g. 
in \cite{GMS} the Lorentz transformations are described as 
'motions in $H_{3}$', in \cite{gilmore} $H_{3}$ is represented 
as a coset space SO(3,1)/SO(3), and last not least there is 
beautiful work \cite{smoro:1963} discussing some physical 
patterns originating from hyperbolic geometry of $H_{3}$. It 
is also possible to delve into further aspects of points and 
lines in $H_{3}$ or projective geometry in $\mathbbm{R}^{4}$ or
$\mathbbm{R}^{5}$, however, here we want to emphasize that this 
SO(3,1) symmetry occurs in the same context as the SO(2,1) 
symmetry, i.e. both groups act as isometries on the respective 
subspaces $H_{2}$ and $H_{3}$ within the coset space $H_{5}$. 
The coset space itself, however, occured because of our procedure 
of building USp(4) equivalence classes, i.e. we have imposed a 
certain (quadratic) constraint on the original quaternionic
coordinates (see above). So for now we introduce the subscript
$\delta_{LG}\,\cdot\, = [Q_{j0}, \cdot\,]$. Regarding Goldstone 
bosons within the quantum point of view, the photon(s) (and further 
'gauge bosons') play(s) the exceptional r\^{o}le as we have a common 
vacuum structure in $H_{5}$ (also for the boundary of $H_{2}$ versus 
$H_{3}$), and we know that Bremsstrahlung occurs whenever we 
accelerate charged particles, i.e.~when we change the equivalence 
classes and the 'vacuum definition', respectively. So it is 
intuitively quite plain (although for now a conjecture), that 
the photon in classical physics plays a residual-only r\^{o}le, 
however, motivated by geometry. More qualitative and quantitative 
aspects have to be worked out profoundly by coordinate mappings
in $\mathbbm{R}^{1,n}$.

Here, the identification of 'classical' coordinates is still open. 
We parametrize elements in $H_{5}$ by five real coordinates of the 
hypercomplex operator set $\{iQ_{01}, iQ_{03}, Q_{j2}\}$ which we 
map via $\exp$ to coordinates of the coset. There, we can introduce 
{\it six} real coordinates which fulfil an hyperbolic constraint 
with signature $(1,5)$, i.e.~by an appropriate involution of the 
hypercomplex operators it is possible to define a purely real 'norm' 
with signature $(1,5)$ which corresponds to an appropriate metric. 
Hence we can understand the hypercomplex description as {\it roots} 
of a real theory (in the sense of Dirac's use of $\gamma$-matrices 
for SO(3,1)) based on 'vectors' obeying an SO(5,1) symmetry, or 
if we remember the su$*$(4) origin of these 'hypercomplex numbers', 
we can think of a microscopic vs. a 'classical' representation. If
we restrict the coset elements to $H_{2}$ and $H_{3}$, then
according to the su$*$(4) multiplication table we can realize 3- 
and 4-dimensional real spaces with signatures $(1,2)$ and $(1,3)$, 
respectively.

Last not least, we can apply actions of the remaining 6-dimensional 
usp(4) generator set $\{iQ_{j1},iQ_{k3}\}$ to $H_{5}$. This time, 
we see a transition of elements in $H_{2}\longrightarrow H_{3}$ 
and vice versa from $H_{3}\longrightarrow H_{2}$ according to
$\delta_{SC}iQ_{01}=-2b_{k}Q_{k2}\subset H_{3}$, 
$\delta_{SC}iQ_{03}=+2a_{k}Q_{k2}\subset H_{3}$,
$\delta_{SC}Q_{j2}=2\left(b_{j}iQ_{01}-a_{j}iQ_{03}\right)\subset H_{2}$,
if we define $\delta_{SC}\,\cdot\, = [a_{k}\,iQ_{k1}+b_{k}\,iQ_{k3},\cdot\,]$.
So this transformation couples the subspaces $H_{2}$ and $H_{3}$ 
within $H_{5}$, and we expect these equations to identify later on 
charge, mass, $\hbar$ (and combinations thereof) geometrically\footnote{An 
explicit/overall representation theory with a thorough identification 
of physical parameters is ongoing but not yet finished. So I invite 
everybody interested in performing calculations to accelerate this 
interesting process. However, in analogy to $\vec{v}$ and $c$ related 
to $H_{3}$ and Lorentz transformations, it is evident to introduce 
constants and transformation parameters for $H_{2}$, $H_{5}$ and 
their isometries as well.} when comparing to experiment or special 
relativity.

\begin{figure}[ht]
\includegraphics[scale=0.3]{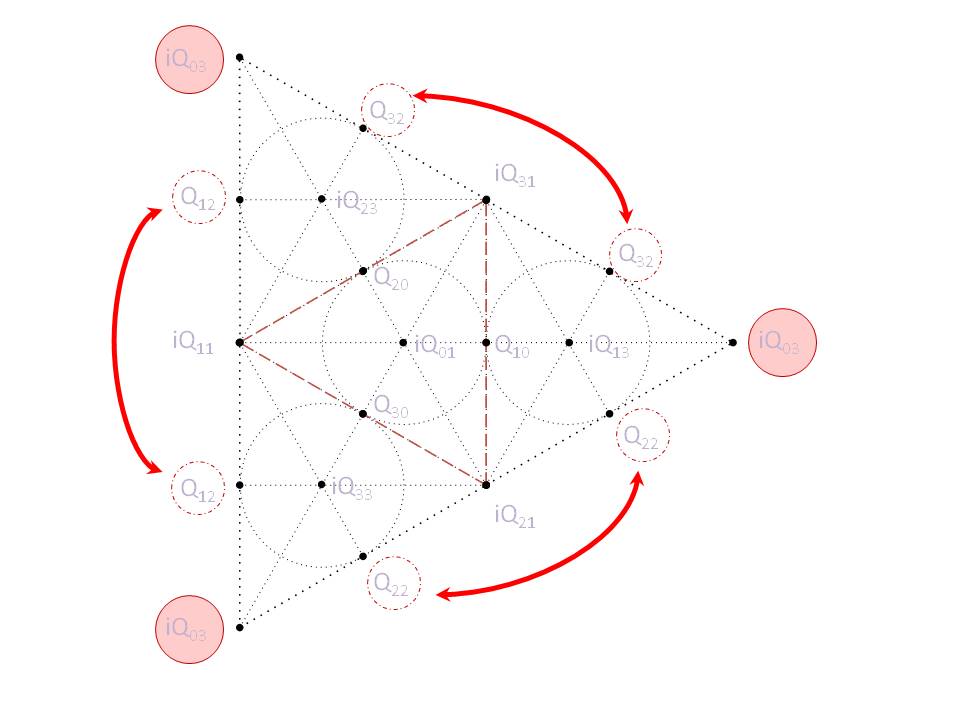}
\includegraphics[scale=0.5]{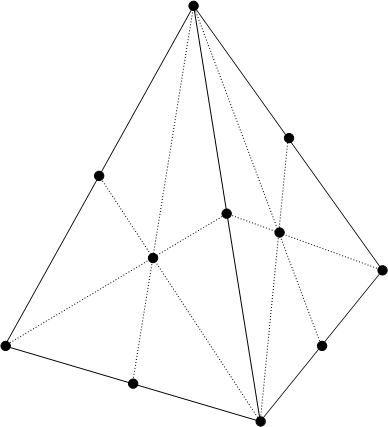}
\caption{Multiplicative structure of the su$*$(4) generators where
identification of the generators leads to a tetrahedron with $iQ_{03}$
on top.}
\label{fig:tetrahedron}
\end{figure}

With respect to the operator algebra itself we can extend the 
graphical representation given in \cite{dahm:2008} by glueing 
four (multiplicative) triangles together (see figure 
\ref{fig:tetrahedron}). In identifying further points at the
edges of the large triangle, we can construct a tetrahedron
where $Q_{02}$ builds not only the U(1) barycenter of the 
tetrahedron but is also involved in spatial multiplications 
of the generator sets, i.e. the full tetrahedral symmetry.

\subsection{Summary and Outlook}
We've presented an end-to-end identification of operators and
symmetries with origin in the low-energy hadron spectrum and
spin$\times$isospin SU(4). We've related this phenomenological
symmetry by the assumption that SU(4) is an approximation of
relativistic SU$*$(4) in the sense that the USp(4) cosets of
SU$*$(4) are responsible for the differences between static 
low-energy and relativistic symmetries, and we've thus studied 
the coset decomposition SU$*$(4)/USp(4) (which differs from 
SU(4)/USp(4) only by complexification ('duality') \cite{gilmore}). 
We've found two kinds of boosts, one related to SO(2,1) and the 
other one related to SO(3,1) so we find our initial assumption 
(in that $\exp(p)$ determines the dynamics) as justified. This 
patterns yields some more exciting results with respect to the 
local and global hyperbolic structure of the manifold $\exp(p)$, 
and using Lie triple systems we've used an algebraic (and 
geometrically more interesting) way to determine the isometries 
as further geometrical properties \cite{dahm:2008} without using 
explicit differential operator representations. Nevertheless, we 
can always introduce such (local) representations by introducing 
(affine) coordinates and representing the generators appropriately 
in terms of differential operators, moreover, we can use as well
nonlinear sigma models or the machinery of differential geometry,
fibers and differential equations, see e.g. the Laplace-Beltrami 
operator in eq.~(\ref{eq:LBop}) or appropriate (differential) 
representations of su$*$(4) operators resp. 'equations of motion'.
Anyhow, we act (locally) with a 10-dimensional compact group on
a noncompact space (remember $[h,p]\subset p$) which controls the 
'time' development of the system.

However, we've emphasized right from the beginning the background
in quaternionic geometry, in that we've represented a noneuclidean 
(projective) geometry in terms of twofold quaternionic coordinates
and related reduction steps. So the very construction of coordinates 
and the restriction and reduction of the coordinates provides a 
continuous background for the various symmetry aspects and patterns 
discussed nowadays in terms of 'separate aspects' or even of 'separate' 
or 'effective' theories. Here, we've used nothing but a 15-dimensional 
operator algebra known for decades in slighty modified/complexified 
form, but we benefit from the enormous power of the division algebra 
$\mathbbm{H}$. We have {\it not} considered additional (unitary or 
gauge) symmetries but the su$*$(4) algebra {\it alone} was not only 
able to provide the known symmetries (local and as isometries), but 
in addition we have seen how the dynamical symmetry breaking breaks 
isospin SU(2) down to an (electromagnetic/electroweak) U(1) symmetry 
in a nontrivial way. Vice versa, strict isospin conservation (or 
in general the assumption of a conserved SU(n) flavour symmetry) 
contradicts the geometrical pattern presented above. So one should
{\it ab initio} expect identification problems when attaching SU(n) 
symmetries to a relativistic description. With respect to an 
experimental verification of our approach the interesting energy 
regime is the {\it low energy-regime} of the particle spectrum 
because the symmetry breaking mechanisms as well as the remaining 
symmetries can be tested in terms of hadron representations and 
their (electromagnetic) interactions \cite{dahm:diss}.

We hope that the use of the hypercomplex representation $Q_{\alpha\beta}$ 
to represent quaternionic M{\"o}bius transformations attends to 
the bell in order to gain more insight into the geometry. Although 
there is still a lot of work to do, the SU$*$(4) representation 
{\it alone} has provided an overall and unique identification of 
well-known unitary symmetries as well as access to their origin(s). 
Moreover, if we understand the hypercomplex system as a description 
of quantum theories (which is justified by its isomorphism to the 
Dirac algebra), the five 'numbers' of the coset basis can be 
interpreted as roots of Lorentz(-like) vectors obeying (orthogonal) 
'norms' with signature(s) $(1,n)$, moreover they are related to the
Dirac equation\cite{dahm:2001}. Thus QPT yields structural and physical 
results based on nothing but a welldefined and transparent geometrical 
foundation. The microscopic structure of spacetime events in $H_{3}$ 
as given by $\{Q_{j2}\}$ corresponds to the Pauli representation 
$iq_{j}\sim \sigma_{j}$ of spacetime events in Sl(2, $\mathbbm{C}$)
\cite{dahm:2006}, however, that way the noncommuting operators 
$Q_{j2}=Q_{02}Q_{j0}$ are only 'approximated'. Last not least, an 
association of quaternions with spacetime events via $\{Q_{j2}\}$ 
has some more interesting consequences for gravitational models as 
well as for (quantum) statistics and for the motivation of (spin) 
lattices and nets, but that's beyond the scope of discussion here.

\end{document}